# New Results for Neutron Radiative Capture on $^{10}$Be at Energies between 25.3 meV and 10.0 MeV

Dubovichenko S.B.[1,2,*], Burkova N.A.[2], Afanasyeva N.V.[1,2], Dzhazairov-Kakhramanov A.V.[1,*], Tkachenko A.S.[1,2]

[1] Fesenkov Astrophysical Institute "NCSRT" ASA MDASI RK, 050020, Almaty, RK

[2] al-Farabi Kazakh National University, RK, 050040, Almaty, RK

**ABSTRACT:** Using the framework of the modified potential cluster model, we succeed in correctly describing the available experimental data for neutron radiative capture on $^{10}$Be total cross sections at low, astrophysical and thermal energies. Unlike our earlier work, the present calculations are based on new experimental data for Coulomb dissociation provided by Prof. T. Aumann and Prof. T. Nakamura. The energy range was extended from $10^{-5}$ to $10^4$ keV for the theoretical cross sections, covering a range of temperatures between 0.01 and 10 $T_9$. The role of the halo asymptotics of the extra-core neutron in $^{11}$Be was also taken into account. The parametrization of the reaction rates for the processes $^{10}$Be$(n,\gamma_{0+1})^{11}$Be are obtained in an analytical form that is convenient for future calculations of different scenarios involving element synthesis in *r*-processes, as widely discussed in the context of *boron* and *beryllium* chains in our previous work.

## 1. Introduction

The main purpose of our research is to bring reliable input to the "round-table" discussion of the relative *boron-beryllium* branching, leading to the production of *carbon components* in astrophysical processes and terrestrial reactor constructive problems which may overlap in some temperature ranges. We are motivated by a set of excellent papers and reviews cited here that directly cover the *pre-history* of these questions.

*1.1. Astrophysical Aspects*

To continue studying the radiative capture processes [1-3], let us consider the $n+^{10}$Be$\rightarrow^{11}$Be$+\gamma$ reaction at low, astrophysical, and thermal energies within the framework of the modified potential cluster model (MPCM) [1] with forbidden states (FSs) [2].

This reaction is part of one of the chains of primordial nucleosynthesis in the early Universe [4]:

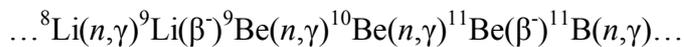
…$^{8}$Li$(n,\gamma)^{9}$Li$(\beta^-)^{9}$Be$(n,\gamma)^{10}$Be$(n,\gamma)^{11}$Be$(\beta^-)^{11}$B$(n,\gamma)$…

since elements with $A > 11$–$12$ can be synthesized (see, for example, [5]).

Alternative scenarios involving fusion reactions in supernovae and the early universe have been suggested in excellent conceptual work [6] in which two possible scenarios for *r*-processes, conditional on $\alpha$-induced reactions and competitive $(n,\gamma)$ neutron radiative capture processes, were analyzed in detail. $(n,\gamma)$ channels possibility justification was given, i.e. when $^{10}$Be$(\alpha,\gamma)^{14}$C reaction is excluded, the

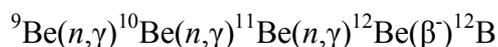
$^{9}$Be$(n,\gamma)^{10}$Be$(n,\gamma)^{11}$Be$(n,\gamma)^{12}$Be$(\beta^-)^{12}$B

---

[*] Corresponding author

reaction chain is started.

The authors of [6] concluded that taking into account the reactions with light neutron excess nuclei could change the estimation of the abundance of heavy elements by up to factor of 10.

However, it is clear that we need reliable reaction rate data for many reactions, including ($n,\gamma$) processes, and some of these are given in [1]. The new and more accurate calculations of these reaction rates presented in the current work therefore enable us to verify the conclusions drawn in [6].

Fig. 6 of [6] illustrates the critical boron and carbon isotope fusion reactions, including ($n,\gamma$) processes. In [1], we considered all of the ($n,\gamma$) reactions presented in this figure except for part of the boron chain from $n^{12}$B, which we intend to address in the future.

Another aspect has appeared in prior works [7,8] devoted to fusion reactions in supernovae, in which the chain of reactions with *boron* isotopes was highlighted. This chain leads to the initiation of synthesis of one of the variants of carbon isotopes (up to $^{19}$C). In addition, the calculations presented here for the $n+^{10}$Be → $^{11}$Be+γ process allow the issues formulated in [8] to be clarified. The latter detailed review suggests a comparative analysis of the rates of 18 reactions providing the *boron* path, leading to an abundance of carbon isotopes as the background for the synthesis of heavier elements. In our opinion, the question of the dominant path is still open. At this time, we see no contradiction between the interpretations suggested in [6] and [7,8].

$^{A}$B$(n,\gamma)^{A+1}$B (A = 11,12,13),

*1.2. Physical Aspects*

We have already considered the neutron radiative capture reaction on $^{10}$Be in [9], in which a comparative analysis was performed using data for the total cross sections [10]. These data [10] were obtained from a recalculation of the experimental measurements of the $\frac{dB}{dE_\gamma}$ Coulomb dissociation probability for $^{11}$Be [11].

However, new results for $^{11}$Be Coulomb dissociation have subsequently been obtained, and these are presented in [12,13]. Moreover, the results reported by Nakamura and Kondo in [13] should be seen as an improvement on the previous data presented in [11]. We now have the opportunity to compare the total cross sections from these different sets of experimental data [11-13] with each other and with previous data [10].

A sequential analysis of these different experimental datasets is presented in Section 2. An extended and more accurate calculation of the $n^{10}$Be channel structural characteristics, performed in the framework of the MPCM [5] and agreeing with the new experimental data in Section 2, is given in Section 3. Section 4 presents theoretical calculations of the radiative capture total cross sections, reaction rates, and analytical parametrizations for astrophysical calculations of the $^{11}$Be isotope synthesis balance.

**2. Processing of Experimental Data**

In this section, we present the formalities of the recalculation of Coulomb dissociation data into the radiative capture total cross sections, since we are faced with the problem of several differences in numerical results caused by the use of different initial data by different authors, for instance the binding energy in the $n^{10}$Be channel and $^{11}$Be Coulomb dissociation measuring results.

We consider the total cross section of the *A*(γ,*n*)*B* direct photodisintegration process, in which $B = A-1$ is expressed in terms of the reduced transition probability of Coulomb dissociation (see, for example [13]), and can be written as

$$\sigma_{\gamma n} = \frac{16\pi^3}{9\hbar c} E_\gamma \cdot \frac{dB(E1)}{dE_\gamma}$$

Here, the excitation energy is equal to $E^* \equiv E_\gamma$, $E_{cm} = E_\gamma - E_b$, where $E_b$ is the binding energy of a neutron, and the constant $e^2/\hbar c = 0.007297$ [14]. The neutron radiative capture cross section is connected with the photodisintegration cross section via a specific equilibrium [13]:

$$\sigma_{n\gamma}(E_{c.m.}) = \frac{2J_A + 1}{2J_{A-1} + 1} \cdot \frac{E_\gamma^2}{2\mu c^2 E_{c.m.}} \cdot \sigma_{\gamma n}(E_\gamma),$$

where $\mu c^2$ is the reduced mass in MeV, which is 853.590 MeV for the reaction under consideration at $m(^{10}Be) = 9327.546$ MeV [15] and $m_n = 939.573$ MeV, $J_A = 1/2$, $J_{A-1} = 0$, and $E_b = 0.507$ MeV [12] or 0.504 MeV [11,13]. It is obvious that the value of the binding energy has a strong influence on the energy of $E_{cm}$ particles at low energies, and various values of this energy have been used in previous works.

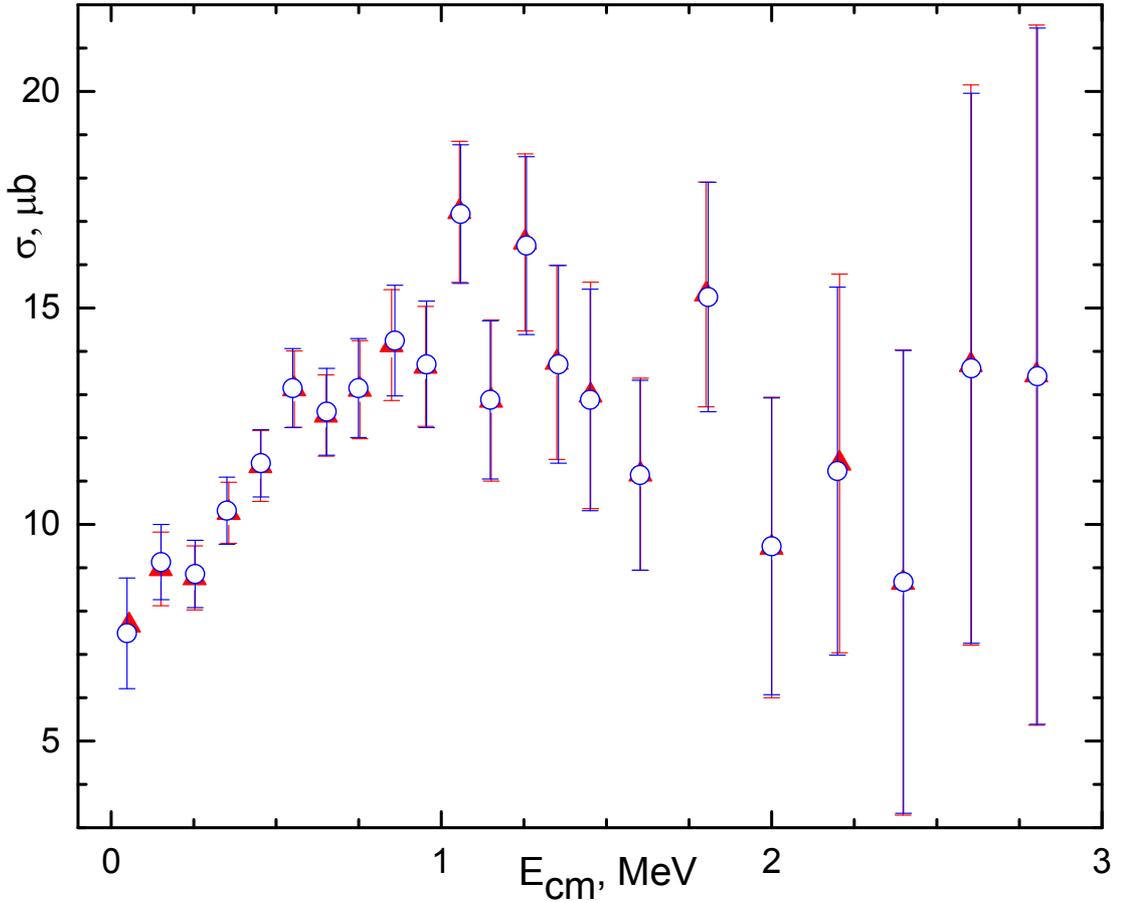

**Fig. 1.** Comparison of the total capture cross sections for $^{14}C(n,\gamma)^{15}C$ obtained using Coulomb dissociation data (red triangles represent the present calculation) and the results for these cross sections (white points) presented in [13]

**Table 1.** Reduced probabilities of Coulomb dissociation, and total cross sections of $^{11}$Be$(\gamma,n)^{10}$Be photodisintegration and $^{10}$Be$(n,\gamma)^{11}$Be radiative capture. The δ symbol represents the absolute errors in the data, while the first three columns correspond to the data in [16]. The probability is measured in $e^2$fm$^2$/MeV, the total cross section is given in μb, and the energy is given in MeV

| $E_\gamma$ | $dB/dE_\gamma$ | $\delta[dB/dE\gamma]$ | $E_n$ | $\sigma(\gamma,n)$ | $\delta[\sigma(\gamma,n)]$ | $\sigma(n,\gamma)$ | $\delta[\sigma(n,\gamma)]$ |
|---|---|---|---|---|---|---|---|
| 0.52500E+00 | 0.13430E+00 | 0.23950E-01 | 0.19813E-01 | 0.28368E+03 | 0.50589E+02 | 0.50889E+01 | 0.90751E+00 |
| 0.57500E+00 | 0.26770E+00 | 0.32620E-01 | 0.74850E-01 | 0.61931E+03 | 0.75464E+02 | 0.35276E+01 | 0.42985E+00 |
| 0.62500E+00 | 0.41420E+00 | 0.41430E-01 | 0.12989E+00 | 0.10416E+04 | 0.10418E+03 | 0.40393E+01 | 0.40403E+00 |
| 0.67500E+00 | 0.36680E+00 | 0.44020E-01 | 0.18492E+00 | 0.99615E+03 | 0.11955E+03 | 0.31650E+01 | 0.37983E+00 |
| 0.72500E+00 | 0.45310E+00 | 0.49170E-01 | 0.23996E+00 | 0.13217E+04 | 0.14343E+03 | 0.37333E+01 | 0.40513E+00 |
| 0.77500E+00 | 0.48750E+00 | 0.48250E-01 | 0.29500E+00 | 0.15201E+04 | 0.15045E+03 | 0.39910E+01 | 0.39501E+00 |
| 0.82500E+00 | 0.57460E+00 | 0.50710E-01 | 0.35003E+00 | 0.19073E+04 | 0.16832E+03 | 0.47824E+01 | 0.42206E+00 |
| 0.87500E+00 | 0.48410E+00 | 0.51430E-01 | 0.40507E+00 | 0.17043E+04 | 0.18106E+03 | 0.41539E+01 | 0.44130E+00 |
| 0.95000E+00 | 0.59000E+00 | 0.37860E-01 | 0.48762E+00 | 0.22551E+04 | 0.14471E+03 | 0.53822E+01 | 0.34537E+00 |
| 0.10500E+01 | 0.57530E+00 | 0.38790E-01 | 0.59770E+00 | 0.24304E+04 | 0.16387E+03 | 0.57810E+01 | 0.38979E+00 |
| 0.11500E+01 | 0.53680E+00 | 0.41580E-01 | 0.70777E+00 | 0.24837E+04 | 0.19239E+03 | 0.59846E+01 | 0.46356E+00 |
| 0.12500E+01 | 0.47300E+00 | 0.39440E-01 | 0.81784E+00 | 0.23788E+04 | 0.19835E+03 | 0.58606E+01 | 0.48867E+00 |
| 0.13500E+01 | 0.40120E+00 | 0.38150E-01 | 0.92792E+00 | 0.21791E+04 | 0.20721E+03 | 0.55192E+01 | 0.52482E+00 |
| 0.14500E+01 | 0.37890E+00 | 0.39470E-01 | 0.10380E+01 | 0.22105E+04 | 0.23026E+03 | 0.57738E+01 | 0.60145E+00 |
| 0.16000E+01 | 0.32110E+00 | 0.29730E-01 | 0.12031E+01 | 0.20670E+04 | 0.19138E+03 | 0.56718E+01 | 0.52514E+00 |
| 0.18000E+01 | 0.27860E+00 | 0.27970E-01 | 0.14232E+01 | 0.20176E+04 | 0.20256E+03 | 0.59230E+01 | 0.59464E+00 |
| 0.20000E+01 | 0.25270E+00 | 0.28770E-01 | 0.16434E+01 | 0.20334E+04 | 0.23150E+03 | 0.63823E+01 | 0.72663E+00 |
| 0.22000E+01 | 0.20650E+00 | 0.29800E-01 | 0.18635E+01 | 0.18278E+04 | 0.26377E+03 | 0.61217E+01 | 0.88342E+00 |
| 0.24000E+01 | 0.17000E+00 | 0.29390E-01 | 0.20837E+01 | 0.16415E+04 | 0.28379E+03 | 0.58516E+01 | 0.10116E+01 |
| 0.26000E+01 | 0.14130E+00 | 0.25010E-01 | 0.23038E+01 | 0.14781E+04 | 0.26162E+03 | 0.55929E+01 | 0.98993E+00 |
| 0.28000E+01 | 0.10960E+00 | 0.27320E-01 | 0.25240E+01 | 0.12347E+04 | 0.30777E+03 | 0.49456E+01 | 0.12328E+01 |
| 0.30000E+01 | 0.12030E+00 | 0.26590E-01 | 0.27441E+01 | 0.14520E+04 | 0.32094E+03 | 0.61411E+01 | 0.13574E+01 |
| 0.32000E+01 | 0.65070E-01 | 0.26880E-01 | 0.29643E+01 | 0.83776E+03 | 0.34607E+03 | 0.37319E+01 | 0.15416E+01 |
| 0.34000E+01 | 0.95910E-01 | 0.26130E-01 | 0.31844E+01 | 0.13120E+04 | 0.35744E+03 | 0.61418E+01 | 0.16733E+01 |
| 0.36000E+01 | 0.30730E-01 | 0.28190E-01 | 0.34046E+01 | 0.44510E+03 | 0.40831E+03 | 0.21849E+01 | 0.20043E+01 |
| 0.38000E+01 | 0.34140E-01 | 0.23230E-01 | 0.36247E+01 | 0.52196E+03 | 0.35516E+03 | 0.26814E+01 | 0.18245E+01 |
| 0.40000E+01 | 0.45360E-01 | 0.23380E-01 | 0.38449E+01 | 0.73000E+03 | 0.37627E+03 | 0.39174E+01 | 0.20191E+01 |
| 0.42000E+01 | 0.52790E-01 | 0.24560E-01 | 0.40650E+01 | 0.89205E+03 | 0.41502E+03 | 0.49918E+01 | 0.23224E+01 |
| 0.44000E+01 | 0.64780E-01 | 0.22730E-01 | 0.42851E+01 | 0.11468E+04 | 0.40239E+03 | 0.66812E+01 | 0.23443E+01 |
| 0.46000E+01 | 0.12210E-01 | 0.23220E-01 | 0.45053E+01 | 0.22598E+03 | 0.42974E+03 | 0.13686E+01 | 0.26028E+01 |
| 0.48000E+01 | -0.99970E-02 | 0.20930E-01 | 0.47254E+01 | -0.19306E+03 | 0.40420E+03 | -0.12139E+01 | 0.25414E+01 |
| 0.50000E+01 | 0.25010E-01 | 0.21770E-01 | 0.49456E+01 | 0.50312E+03 | 0.43794E+03 | 0.32797E+01 | 0.28548E+01 |
| 0.52000E+01 | 0.15750E-01 | 0.18990E-01 | 0.51657E+01 | 0.32951E+03 | 0.39730E+03 | 0.22242E+01 | 0.26818E+01 |

In order to test the calculation methods and computer programs used here, we consider the $^{14}$C$(n,\gamma)^{15}$C reaction detailed in [13]. Fig. 2.10 of this work presents data on the Coulomb dissociation probability, and Fig. 2.11 shows the results of a recalculation into the total cross-sections of the neutron radiative capture on $^{10}$Be. We have digitized the data presented in Fig. 2.10 and have recalculated these into the total cross sections. The results of the present recalculation and data from Fig. 2.11 of [13] are presented in Fig. 1. In this recalculation, we used a mass $m(^{14}C) = 13{,}043.936$ MeV, and the neutron mass given above. As can be seen from Fig. 1, the difference between the present results and the data in [13] for the neutron capture on $^{14}$C total cross sections is small; this arises from the inaccuracy in the digitization of data for the Coulomb dissociation probability from the figures presented in [13].

The following is a recalculation of data for the $^{11}$Be Coulomb dissociation probability from [11-13], as shown in Fig. 2, into $^{10}$Be$(n,\gamma)^{11}$Be radiative capture total cross sections. The results of this recalculation are shown in Fig. 3, while those for the Coulomb dissociation data [12] provided by Prof. T. Aumann [16] are given in Table 1.

A similar recalculation of the data for the $^{10}$Be$(n,\gamma)^{11}$Be reaction provided by Prof. T. Nakamura [13] is presented in Table 2.

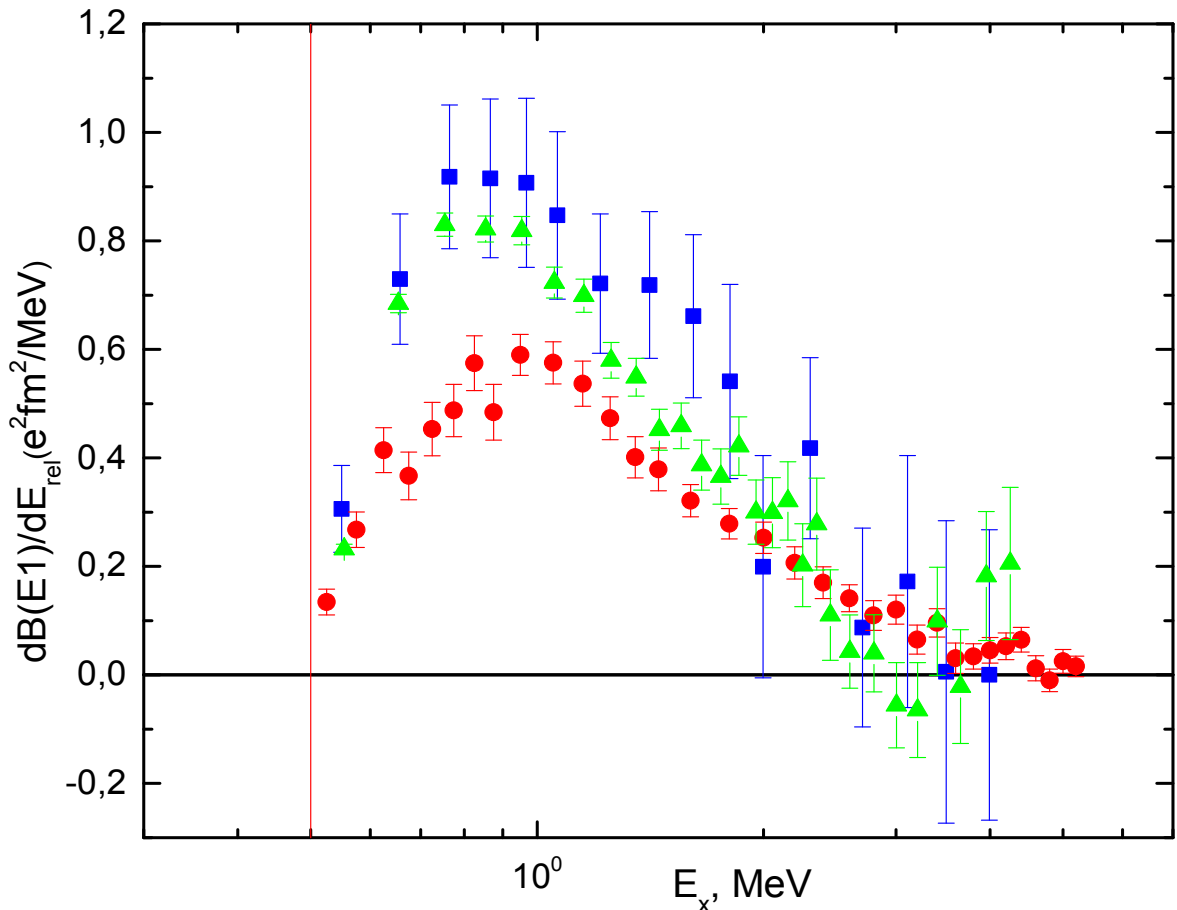

**Fig. 2.** Data on Coulomb dissociation probabilities from [11-13]: blue squares [11], red points [12], and green triangles [13]

In Fig. 2, the vertical red line indicates an energy of 0.5 MeV. The red points illustrate the results from [12], blue squares correspond to the data from [11], and green triangles indicate the results from [13]. It can be seen from Fig. 2 that there is a substantial difference between the data from [11,13] and those from [12]; the latter are substantially lower than the former, and this difference can reach 50%.

**Table 2.** Reduced probabilities of Coulomb dissociation, and total cross sections for $^{11}$Be$(\gamma,n)^{10}$Be photodisintegration and $^{10}$Be$(n,\gamma)^{11}$Be radiative capture. The δ symbol shows the absolute errors in the data, and the first three columns correspond to the data in [13].

| $E_\gamma$ | $dB/dE_\gamma$ | $\delta[dB/dE_\gamma]$ | $E_n$ | $\sigma(\gamma,n)$ | $\delta[\sigma(\gamma,n)]$ | $\sigma(n,\gamma)$ | $\delta[\sigma(n,\gamma)]$ |
|---|---|---|---|---|---|---|---|
| 0.55400E+00 | 0.23220E+00 | 0.88112E-02 | 0.55037E-01 | 0.51755E+03 | 0.19640E+02 | 0.37218E+01 | 0.14123E+00 |
| 0.65400E+00 | 0.68446E+00 | 0.16999E-01 | 0.16511E+00 | 0.18010E+04 | 0.44728E+02 | 0.60163E+01 | 0.14942E+00 |
| 0.75400E+00 | 0.82996E+00 | 0.21356E-01 | 0.27518E+00 | 0.25178E+04 | 0.64787E+02 | 0.67077E+01 | 0.17260E+00 |
| 0.85400E+00 | 0.82182E+00 | 0.23971E-01 | 0.38526E+00 | 0.28238E+04 | 0.82363E+02 | 0.68933E+01 | 0.20106E+00 |
| 0.95400E+00 | 0.81894E+00 | 0.26301E-01 | 0.49533E+00 | 0.31433E+04 | 0.10095E+03 | 0.74478E+01 | 0.23919E+00 |
| 0.10540E+01 | 0.72331E+00 | 0.28438E-01 | 0.60540E+00 | 0.30673E+04 | 0.12060E+03 | 0.72581E+01 | 0.28537E+00 |
| 0.11540E+01 | 0.69900E+00 | 0.30684E-01 | 0.71548E+00 | 0.32454E+04 | 0.14247E+03 | 0.77897E+01 | 0.34195E+00 |
| 0.12540E+01 | 0.57986E+00 | 0.32868E-01 | 0.82555E+00 | 0.29256E+04 | 0.16583E+03 | 0.71861E+01 | 0.40733E+00 |
| 0.13540E+01 | 0.54881E+00 | 0.34808E-01 | 0.93562E+00 | 0.29897E+04 | 0.18962E+03 | 0.75543E+01 | 0.47914E+00 |
| 0.14540E+01 | 0.45194E+00 | 0.37923E-01 | 0.10457E+01 | 0.26438E+04 | 0.22185E+03 | 0.68927E+01 | 0.57838E+00 |
| 0.15540E+01 | 0.45915E+00 | 0.41971E-01 | 0.11558E+01 | 0.28708E+04 | 0.26242E+03 | 0.77350E+01 | 0.70705E+00 |
| 0.16540E+01 | 0.38684E+00 | 0.46088E-01 | 0.12658E+01 | 0.25743E+04 | 0.30670E+03 | 0.71743E+01 | 0.85475E+00 |
| 0.17540E+01 | 0.36575E+00 | 0.51044E-01 | 0.13759E+01 | 0.25811E+04 | 0.36022E+03 | 0.74422E+01 | 0.10386E+01 |
| 0.18540E+01 | 0.42178E+00 | 0.54085E-01 | 0.14860E+01 | 0.31462E+04 | 0.40344E+03 | 0.93847E+01 | 0.12034E+01 |
| 0.19540E+01 | 0.30008E+00 | 0.59193E-01 | 0.15961E+01 | 0.23591E+04 | 0.46536E+03 | 0.72775E+01 | 0.14356E+01 |
| 0.20540E+01 | 0.29910E+00 | 0.64603E-01 | 0.17061E+01 | 0.24718E+04 | 0.53388E+03 | 0.78819E+01 | 0.17024E+01 |
| 0.21540E+01 | 0.32079E+00 | 0.72091E-01 | 0.18162E+01 | 0.27801E+04 | 0.62476E+03 | 0.91584E+01 | 0.20581E+01 |
| 0.22540E+01 | 0.20229E+00 | 0.76464E-01 | 0.19263E+01 | 0.18345E+04 | 0.69343E+03 | 0.62392E+01 | 0.23584E+01 |
| 0.23540E+01 | 0.27806E+00 | 0.84503E-01 | 0.20364E+01 | 0.26335E+04 | 0.80033E+03 | 0.92412E+01 | 0.28084E+01 |
| 0.24540E+01 | 0.11028E+00 | 0.83553E-01 | 0.21464E+01 | 0.10889E+04 | 0.82495E+03 | 0.39395E+01 | 0.29846E+01 |
| 0.26040E+01 | 0.42889E-01 | 0.67467E-01 | 0.23115E+01 | 0.44935E+03 | 0.70685E+03 | 0.16998E+01 | 0.26739E+01 |
| 0.28040E+01 | 0.39987E-01 | 0.71057E-01 | 0.25317E+01 | 0.45111E+03 | 0.80163E+03 | 0.18066E+01 | 0.32103E+01 |
| 0.30040E+01 | -0.56019E-01 | 0.78597E-01 | 0.27518E+01 | -0.67706E+03 | 0.94994E+03 | -0.28631E+01 | 0.40171E+01 |
| 0.32040E+01 | -0.64728E-01 | 0.87340E-01 | 0.29720E+01 | -0.83441E+03 | 0.11259E+04 | -0.37166E+01 | 0.50150E+01 |
| 0.34040E+01 | 0.98931E-01 | 0.99400E-01 | 0.31921E+01 | 0.13549E+04 | 0.13613E+04 | 0.63423E+01 | 0.63723E+01 |
| 0.36540E+01 | -0.21345E-01 | 0.10489E+00 | 0.34673E+01 | -0.31380E+03 | 0.15420E+04 | -0.15582E+01 | 0.76568E+01 |
| 0.39540E+01 | 0.18232E+00 | 0.11905E+00 | 0.37975E+01 | 0.29005E+04 | 0.18939E+04 | 0.15398E+02 | 0.10055E+02 |
| 0.42540E+01 | 0.20553E+00 | 0.14036E+00 | 0.41277E+01 | 0.35177E+04 | 0.24023E+04 | 0.19887E+02 | 0.13581E+02 |

In Fig. 3, the black points show the results of recalculation of the data in [11], performed earlier in [10] and used by the current authors in [9]. It can be seen that these differ substantially from the present recalculation of the same data [11], shown in Fig. 3 using blue squares.

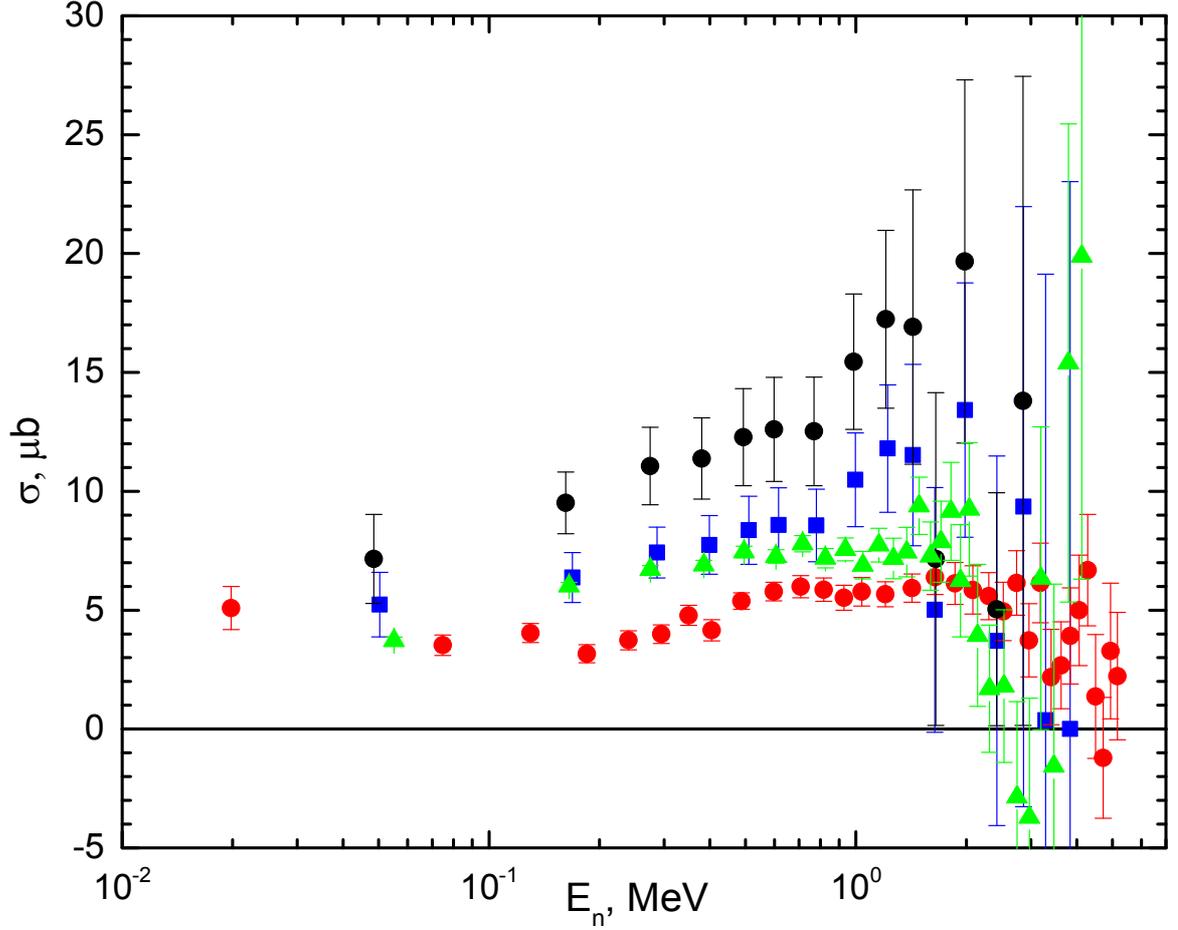

**Fig. 3.** Data obtained by the current authors for $^{10}$Be$(n,\gamma)^{11}$Be total radiative capture cross sections, based on the Coulomb dissociation probability data in [11-13]: blue squares [11], red points [12], and green triangles [13]. Black points show the results of recalculation of the data in [11], performed previously in [10].

## 3. Model and Methods

Let us consider the basic methods of calculating the total radiative capture cross-sections, the cluster state classifications according to Young tableaux, and the principles for construction of the intercluster interaction in the MPCM used here.

*3.1. General Definitions*

The total cross sections of radiative capture σ($NJ,J_f$) for the *EJ* and *MJ* transitions in the potential cluster model are presented, for example, in [1,3] and [17], and have the form

$$\sigma_c(NJ,J_f) = \frac{8\pi K e^2}{\hbar^2 q^3} \frac{\mu \cdot A_J^2(NJ,K)}{(2S_1+1)(2S_2+1)} \frac{J+1}{J[(2J+1)!!]^2} \sum_{L_i,J_i} P_J^2(NJ,J_f,J_i) I_J^2(J_f,J_i),$$

where σ is the total radiative capture cross section, μ is the reduced mass of the initial channel particles in amu, *q* is the wave number of initial channel particles, $S_1$, $S_2$ are the initial channel particle spins, *K* and *J* are the wave number and angular momentum of the γ-quantum in the final channel, respectively, and *N* corresponds to the *E* or *M* transitions of *J* multipolarity from the initial state $J_i$ to the final nucleus state $J_f$.

For *EJ*(*L*) electric orbital transitions ($S_i = S_f = S$), the quantity $P_J$ has the form [1,3]

$$P_J^2(EJ, J_f, J_i) = \delta_{S_i S_f}[(2J+1)(2L_i+1)(2J_i+1)(2J_f+1)](L_i 0 J 0 | L_f 0)^2 \begin{Bmatrix} L_i & S & J_i \\ J_f & J & L_f \end{Bmatrix}^2$$

$$A_J(EJ, K) = K^J \mu^J \left( \frac{Z_1}{m_1^J} + (-1)^J \frac{Z_2}{m_2^J} \right), \qquad I_J(J_f, J_i) = \langle \chi_f | r^J | \chi_i \rangle$$

Here, $S_i$, $S_f$, $L_f$, $L_i$, $J_f$ and $J_i$ are the total spins and the momenta of the initial (*i*) and final (*f*) channel particles; $m_1$, $m_2$, $Z_1$ and $Z_2$ are the masses (in amu) and charges (in "*e*" units) of the initial channel particles; $I_J$ is the integral of the wave functions of the initial state $\chi_i$ and the final state $\chi_f$, as functions of the relative motion of clusters with intercluster distance *r*.

The following expression was obtained for the spin part of the *M*1(*S*) magnetic process at *J* = 1 in the model used here ($S_i = S_f = S$, $L_i = L_f = L$) [1,3]:

$$P_1^2(M1, J_f, J_i) = \delta_{S_i S_f} \delta_{L_i L_f} \left[ S(S+1)(2S+1)(2J_i+1)(2J_f+1) \right] \begin{Bmatrix} S & L & J_i \\ J_f & 1 & S \end{Bmatrix}^2 =$$

$$= \delta_{S_i S_f} \delta_{L_i L_f} \left[ S(S+1)(2S+1)(2J_i+1)(2J_f+1) \right] \begin{Bmatrix} L & S & J_i \\ 1 & J_f & S \end{Bmatrix}^2,$$

$$A_1(M1, K) = i \frac{\hbar K}{m_0 c} \sqrt{3} \left[ \mu_1 \frac{m_2}{m} - \mu_2 \frac{m_1}{m} \right], \quad I_J(J_f, J_i) = \langle \chi_f | r^{J-1} | \chi_i \rangle.$$

where *m* is the nucleus mass in amu; $\mu_1$, $\mu_2$ are the cluster magnetic moments; and the remaining notation is the same as in the previous expression. The constant $\hbar^2 / m_0$ is equal to 41.4686 MeV·fm$^2$.

*3.2. Principles for the Construction of Interaction Potentials*

Previously, using the framework of the modified potential cluster model with forbidden states, we demonstrated the possibility of describing the astrophysical S-factors of radiative capture reactions on many light atomic nuclei [1,3]. This model takes into account the supermultiplet symmetry of the cluster system wave function (WF) with separation of orbital states using Young tableaux. The orbital state classification enables us to analyze the structure of intercluster interactions, and to determine the presence and quantity of the allowed states (ASs) and FSs in the intercluster wave functions, thus giving us an opportunity to find the number of nodes of the relative motion WF of the cluster. For any cluster system, the task many-particle character and antisymmetrization effects are qualitatively taken into account via the separation of one-particle bound levels with these potentials into states that are allowed or forbidden by the Pauli exclusion principle [2].

In the approach used here, the intercluster interaction potentials for scattering processes are constructed on the basis of a description of the elastic scattering phase shifts, taking into account their resonance behavior. These phase shifts arise from the experimental data for differential cross sections by applying phase shift analysis. For the bound states (BSs) of light nuclei in the cluster channels, the potentials are constructed not only on the basis of the scattering phase shifts description but also by using certain additional requirements. For example, one of these requirements is the reproduction of binding energy and several other basic characteristics of the nuclei bound states, and in some cases this requirement is essential. In addition, it is assumed that a bound nuclear ground state is relevant to the cluster channel, which consists of initial particles participating in the capture reaction [18,19].

The choice of the modified potential cluster model for the description of cluster systems in nuclei and in nuclear and thermonuclear processes at astrophysical energies [20] is based on the fact that the possibility of forming nucleon associations (i.e. clusters) in many light atomic nuclei and the degree of their isolation from each other are comparatively high. Many experimental measurements and various theoretical calculations by different authors over the last 50 to 60 years have confirmed this [2,21]. This assumption is of course an idealization of the actual situation in the nucleus, since it assumes that there is 100% clusterization of the nucleus in the BS for the initial channel particles.

If one cluster channel dominates in the nucleus structure, then the one-channel cluster model used here allows us to identify this dominant cluster channel and to describe the properties of the nuclear system [22].

We now consider in detail the procedure of construction of the partial intercluster potentials used here, for a given orbital moment $L$ and other quantum numbers, by defining the criteria for and sequence of finding parameters, and by specifying their possible errors and ambiguities. Firstly, the parameters of the ground state (GS) potentials are found. These parameters for a given number of allowed and forbidden states in the partial wave BSs are uniquely fixed by the binding energy, nucleus radius and an asymptotic constant in the considered channel.

The accuracy of this calculation of BS potential parameters is related to the accuracy of the asymptotic constants (ACs), which is equal to 10%–20%. There are no other ambiguities in this potential, because the classification of states according to Young tableaux allows us to uniquely fix the number of the BSs that are forbidden or allowed in the given partial wave. This number completely determines the depth of the potential, while the potential width depends entirely on the AC value. Principles for the numerical determination of the FSs and ASs in the partial wave are presented below.

It should be noted here that calculations of the charge radius in any model contain the model errors, i.e. those arising from the accuracy of the model. In any model, the radius values depend on the integral of the model's WF, i.e. the model errors of such functions can be simply summarized. At the same time, the AC values are determined by the model WFs at one point of their asymptotics, and appear to involve very little error. The BSs potentials constructed here therefore agree with the AC values obtained by independent methods that allow for a determination the AC from the experimental data [23].

The intercluster potential of the nonresonance capture process is also constructed uniquely using the scattering phase shifts for the number of the allowed and forbidden BSs in the partial wave. The determination of the accuracy of these potential parameters mainly depends on the accuracy of determination of scattering phase shifts from the experimental data, and may reach 20%–30%. As in the previous case, this potential has no ambiguities since the classification of states based on Young tableaux allows us to uniquely fix the number of BSs, which completely determines its depth. The width of this potential at its depth is determined by a form of the elastic scattering phase shift.

Upon construction of the nonresonance scattering potential using the data for nuclei spectra in the specified channel, it is difficult to evaluate the accuracy of calculation of its parameters, even for a given number of BSs, though one may seemingly hope that this accuracy does not exceed the error in the previous case. For the energy region up to 1 MeV, it is usually assumed that this potential must lead to a near-zero value of the scattering phase shift, or to a smoothly decreasing form of the phase shift, if there are no resonance levels in the nucleus spectra.

In the case of resonance scattering analysis, the potential can be constructed uniquely, since at the given BSs number and energies up to 1 MeV there is a comparatively narrow resonance in the considered partial wave, the width of that is of the order of 10÷50 keV. For a given number of BSs, the potential depth is fixed uniquely by the resonance level energy, and its width is completely determined by this resonance width. The error in its parameters does not as a rule exceed the error in the resonance width, and is equal to about 3%–5%. Moreover, it is also concerned with the construction of the partial potential using the scattering phase shifts and the determination of its parameters on the basis of the resonance in the nucleus spectra.

As a result, the potentials have no ambiguities that are common to the optical model, and, as we show later, they allow us to describe correctly the total radiative capture cross sections. The potentials of the BSs must correctly describe the known values of the AC, which is related to the asymptotic normalizing coefficient (ANC) determined from the experiment and denoted as $A_{NC}$, by the following expression [23,24]

$$A_{NC}^2 = S_f \cdot C^2,$$

where $S_f$ is the spectroscopic factor, and $C$ is the dimensional asymptotic constant expressed in fm$^{-1/2}$

$$\chi_L(r) = C W_{-\eta L+1/2}(2k_0 r),$$

which is related to the non-dimensional AC $C_w$ [25] used here

$$\chi_L(r) = \sqrt{2k_0} C_w W_{-\eta L+1/2}(2k_0 r)$$

as follows:
$$C = \sqrt{2k_0} C_w$$

In summary, it should be emphasized that in the construction of the partial interaction potentials, it is assumed that they depend not only on the orbital moment $L$, but also on the total spin $S$, and the total moment $J$ of the cluster system. In other words, for the different moments $JLS$ we have different parameter values. Since the $E1$ and $M1$ transitions between the different $^{(2S+1)}L_J$ states in the continuous and discrete spectra are usually considered, the potentials of these states will therefore be different.

In addition, one of the modifications of the model used here is an assumption that the intercluster potentials depend explicitly on the Young tableaux $\{f\}$. In other words, if two tableaux are allowed in the states of a continuous spectrum and only one is allowed in the discrete spectrum, these potentials may have different parameters for the same $JLS$, i.e. within the same partial wave [1,2].

## 3.3. Structure of discrete states in $n^{10}$Be

For $^{10}$Be, the orbital Young tableau {442} is used in the same way as for $^{10}$B [1]; thus, for $n^{10}$Be system we have $\{1\} \times \{442\} = \{542\} + \{443\} + \{4421\}$ [26,27]. The first of the obtained tableaux is compatible with the orbital moment $L = 0,2,3,4$, and is forbidden since five nucleons cannot exist in the $s$-shell. The second tableau is allowed and compatible with the orbital moment $L = 1,2,3,4$, and the third is also allowed and compatible with $L = 1,2,3$ [26].

As noted earlier [1,3], due to the lack of product tables for Young tableaux for 10 and 11 particles, it is impossible to accurately classify the cluster states in the particle system under consideration. However, even an assessment of the orbital symmetries with this accuracy enables us to determine the presence of the FSs in the S- and D-waves and the lack of the FSs for P states. Based on these structures of the FSs and ASs in the different partial waves, we can then construct the intercluster interaction potentials required for the calculation of the total radiative capture cross sections.

Thus, taking into account only the lowest partial waves with orbital moment $L = 0,1,2$, we can say that for the $n^{10}$Be system (since we know that for $^{10}$Be $J^\pi, T = 0^+, 1$ [28]), the allowed state (AS) lies only in the $^2P$ wave potentials, while the $^2S$-, $^2D$-waves have the forbidden state (FS). The state in the $^2S_{1/2}$ wave with the FS corresponds to the $^{11}$Be ground state with $J^\pi, T = 1/2^+, 3/2$ and this corresponds to an $n^{10}$Be system binding energy of $-0.5016$ MeV [29].

We note that the $^2P$ waves correspond to two allowed Young tableaux, {443} and {4421}. This situation appears to be the same as in the N$^2$H or N$^{10}$B systems described in previous work [1,30-32], where the scattering potentials depend on two Young tableaux, and the BS potentials are defined by only one [30,33]. The potentials for one partial wave in the scattering states and discrete spectrum may therefore be different. It is also assumed here that the potential of the $^2P_{1/2}$ bound state (see Fig. 4) corresponds to one tableau {443}, which is the first excited state (FES) of $^{11}$Be at an energy of 0.32004 MeV relative to the GS.

For clearness, we consider that for the discrete spectrum the allowed state in the $^2P_{1/2}$ wave is bound, while for the scattering processes it is unbound. The depth of the potential can therefore be set to zero. We turn our attention again to the fact that these potentials may be different since they correspond to states with different Young tableaux. For the potential of the $^2S_{1/2}$ scattering wave or the discrete spectrum in the $n^{10}$B system, the FSs and ASs are bound.

We now consider the FES that is bound in the $n^{10}$Be channel, and several resonance states of $^{11}$Be [29], which are unbound in the $n^{10}$Be channel and correspond to the resonances in the $n^{10}$Be scattering:

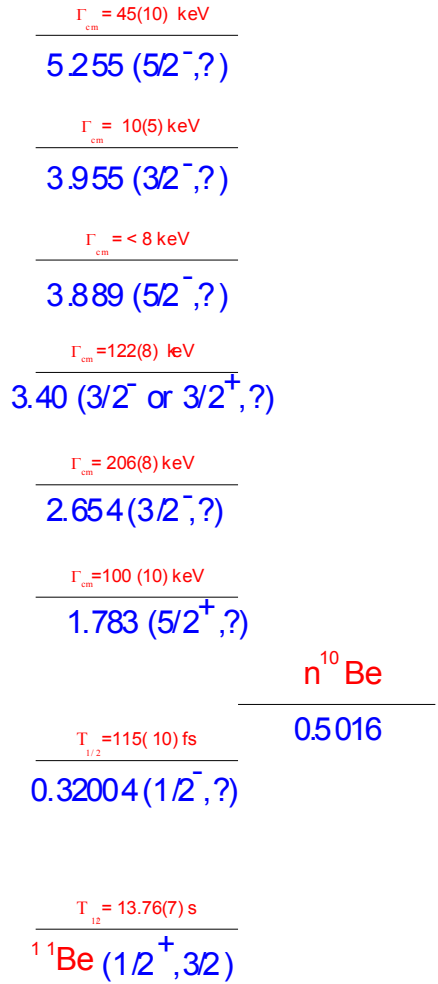

**Fig. 4.** $^{11}$Be energy levels in MeV (c.m.) and their widths.

1. The first excited state (ES) of $^{11}$Be is found at an energy of 0.32004 MeV relative to the GS, with moment $J^\pi = 1/2^-$ or $-0.18156$ MeV relative to the threshold of the $n^{10}$Be channel. This state may be related to the doublet $^2P_{1/2}$ level with the bound AS but without the bound FSs (see Fig. 4).

2. The first resonance state has an excitation energy of 1.783 MeV relative to the GS or 1.2814 MeV (c.m.) relative to the threshold of the $n^{10}$Be channel. For this level, $J^\pi$ is equal to $5/2^+$ [29], leading to $L=2$ and enabling this to be considered as the $^2D_{5/2}$ resonance in the $n^{10}$Be system at 1.41 MeV in the laboratory system (l.s.). The potential at this resonance has a bound FS, and the resonance width is 100(10) keV in the center-of-mass system (c.m.) [29].

3. The second resonance state energy is 2.654 MeV relative to the GS or 2.1524 MeV (c.m.) relative to the channel threshold, with a width of 206(8) keV (c.m.). Its moment $J^\pi$ is $3/2^-$ [29]. These characteristics mean that it can be considered to be $^2P_{3/2}$ resonance in the $n^{10}$Be system at 2.37 MeV (l.s.), and its potential has no bound FSs or ASs.

4. The third resonance state is at an excitation energy of 3.400(6) MeV (c.m.) with a width 122(8) keV (c.m.), and moment $J^\pi = 3/2^+$ (or $3/2^-$) [29]. This leads to $L = 2$ and means that it can be considered the $^2D_{3/2}$ resonance in the $n^{10}$Be system at 3.19 MeV (l.s.). However, the ambiguity in parity of this level enables us to assume that this is also the $^2P_{3/2}$ state;

due to this ambiguity, we do not consider this resonance.

5. We do not consider the subsequent two resonances shown in Fig. 4, because their widths are of the order of 10 keV or less.

6. The resonance at 5.255(3) MeV has a width of 45(10) keV, moment $J^{\pi} = 5/2^-$ and can be correlated with the $^2P_{5/2}$ wave. However, its width is rather smaller than the low resonance width, and therefore we do not consider this.

At higher energies, there are only two resonances. The first of these has an energy of 8.020(20) MeV with a width of 230(55) keV and moment $J^{\pi} = 3/2^-$, while the second has an energy of 10.59(50) MeV, a width of 210(40) keV, and $J^{\pi} = 5/2^-$. These two resonances have fully determined characteristics [29]. The first can be correlated with the $^2P_{3/2}$ state, and in this case the E1 transition to the GS can be considered. However, its width relative to the excitation energy is comparatively small, so the contribution of this potential to the total cross sections at this energy will be relatively small. For the second level, only the M2 transition is possible, and we therefore disregard this. As a result, we take into account only the second resonance at 2.654 MeV, which has a larger width and well-determined quantum numbers; the neutron radiative capture on the $^{10}$Be total cross section will be calculated up to 10 MeV.

*3.4. Classification for capture of the E1 transition*

On the basis of the data given above, it can be assumed that the *E*1 capture of a neutron is allowed from $^2P$ scattering waves (without bound FSs or Ass) to the $^2S_{1/2}$ GS of $^{11}$Be (with bound FSs or Ass).

1. $\begin{array}{l} ^2P_{1/2} \to\, ^2S_{1/2} \\ ^2P_{3/2} \to\, ^2S_{1/2} \end{array}$ .

For radiative capture to the FES, the analogous *E*1 transition is allowed from the $^2S_{1/2}$ and $^2D_{3/2}$ scattering waves with bound FSs but without bound ASs, to the $^2P_{1/2}$ BS with bound AS but without bound FS.

2. $\begin{array}{l} ^2S_{1/2} \to\, ^2P_{1/2} \\ ^2D_{3/2} \to\, ^2P_{1/2} \end{array}$ .

The potentials of the GS and FES will be constructed further in order to correctly describe the channel binding energy, the $^{11}$Be charge radius and its asymptotic constant $C_w$ in the $n^{10}$Be channel. Data for the asymptotic normalizing coefficients $A_{NC}$ are given, for example, in [5]. In our calculations, we use a $^{10}$Be GS radius of 2.357(18) fm, taken from [34], and the $^{11}$Be GS radius value given in [29] of 2.463(15) fm. The charge radius of the neutron is zero and its mass radius is 0.8775(51) fm, which agrees with the known radius of a proton [14].

The estimated value of the charge radius of $^{11}$Be FES presented in [35] is equal to 2.43(10) fm, and the value of $^{11}$Be GS obtained in the same work is equal to 2.42(10) fm. The value of the radius of the neutron in $^{11}$Be given in [35] is 5.6(6) fm. At the same time, the value of the neutron radius in the GS reported in [36] is equal to 7.60(25) fm, and another value for the FES in the same work is given as 4.58(25) fm. In the remaining calculations, we use the exact values of $^{10}$Be and the following neutron masses: $m(^{10}\text{Be}) = 10.01134$ amu [15] and $m_n = 1.008665$ amu [14,15].

## 3.5. Potentials for the $n^{10}Be$ scattering states

As the $n^{10}Be$ interaction in each partial wave with the given orbital moment $L$ and other quantum numbers including $\{f\}$, we usually use the potential in the Gauss-type form with the point-like Coulomb term [1,3] which is simply absent for the neutron processes

$$V(r,JLS\{f\}) = V_0(JLS\{f\})\exp\{-\gamma(JLS\{f\})r^2\}.$$

The GS of $^{11}Be$ in the $n^{10}Be$ channel is the $^2S_{1/2}$ state, and the potential should correctly describe the binding energy and the AC of this channel. To determine this constant from the available experimental data, we examine the data for the spectroscopic factors $S_f$ and asymptotic normalizing coefficients $A_{NC}$, which are connected with the AC. The results for $A_{NC}$ given in [5] are presented in Table 3, to which we add some results from [4].

**Table 3.** Data for $A_{NC}$ of $^{11}Be$ in the $n^{10}Be$ channel

| Reaction from which the $A_{NC}$ was obtained | $A_{NC}$ in fm$^{-1/2}$ for the GS | $A_{NC}$ in fm$^{-1/2}$ for the FES | Reference |
|---|---|---|---|
| $(d,p_0)$ at 12 MeV | 0.723(16) | 0.133(4) | [5] |
| $(d,p_0)$ at 25 MeV | 0.715(35) | 0.128(6) | [5] |
| $(d,p_0)$ at 25 MeV | 0.81(5) | 0.18(1) | [4] |
|  | 0.68–0.86 | 0.122–0.19 | [a)]present |
|  | 0.749 | 0.147 | [b)]present |

[a)] Range  [b)] Average $\overline{A}_{NC}$

We have also found data for the spectroscopic factors of $^{11}Be$ in the $n^{10}Be$ channel [29], and we present these values Table 4 below.

**Table 4.** Data for the spectroscopic $S$-factors of $^{11}Be$ in the $n^{10}Be$ channel

| $S$ for the GS | $S$ for the FES | Reference |
|---|---|---|
| 0.87(13) | --- | [37] |
| 0.72(4) | --- | [38] |
| 0.61(5) | --- | [12] |
| 0.73(6) | 0.6(2) | [39] |
| 0.73(6) | 0.63(15) | [40] |
| 0.77 | 0.96 | [41] |
| 0.56–1.0 | 0.4–0.96 | [a)]present |
| 0.74 | 0.73 | [b)]present |

[a)] Range  [b)] Average $\overline{S}$

On the basis of the expressions given above for the GS, we then obtain the value of $\overline{A}_{NC}/\sqrt{\overline{S}} = \overline{C} = 0.87$ fm$^{-1/2}$. Since $\sqrt{2k_0} = 0.546$, the dimensionless AC determined using $\overline{C}_w = \overline{C}/\sqrt{2k_0}$ is found to be 1.59. However, the range of the spectroscopic $S$-factor values is so wide that the values of $C_w$ for the average ANC may be in the region 1.37–1.83. If we take into account the $A_{NC}$ errors, this range may be extended to 1.25–2.10. Similarly to the GS, for the first excited state at $\sqrt{2k_0} = 0.423$ we obtain $\overline{C}_w = 0.41$, and the $C_w$ values for the average ANC are in the range 0.35 to 0.55. If we take into account the $A_{NC}$ errors, the range of dimensionless AC values may be extended to 0.29–0.71. These results are also used in the construction of the GS and FES potentials, which should agree with the binding energy of these levels and their AC values.

For the specific case of the potential of $^2S_{1/2}$ GS with the FS, the following parameters may be used:

$$V_{S1/2} = 174.15483 \text{ MeV and } \gamma_{1/2} = 0.4 \text{ fm}^{-2}. \tag{1}$$

This leads to a binding energy of $-0.501600$ MeV at an accuracy of $10^{-6}$ MeV for the finite-difference method (FDM) used here [42]. The AC is $C_w = 1.32(1)$ in the range 7–30 fm, the mass radius is 2.88 fm, and the charge radius is 2.43 fm. The AC errors are determined by its averaging over the pointed range of distances.

The potential of the GS with the FS is in total agreement with the classification of states according to the Young tableaux given above, meaning that the $^{11}$Be charge radius is in good agreement with the data [29]. The potential parameters (1) enable us to obtain the AC value at its lower limit. The potential phase shift is shown in Fig. 5 by the red solid line. This potential at an orbital moment $L=2$ leads to the nonresonance $^2D$ scattering phase shift without spin-orbit splitting, as shown in Fig. 5 by the red dotted line. The $n^{10}$Be scattering $^2S_{1/2}$ phase shifts obtained in [43] are shown in the same figure by black points.

For the potential of the resonance $^2D_{5/2}$ wave with the FS, which is used in to calculate the $E2$ transitions, the following parameters are obtained

$$V_{D5/2} = 474.505 \text{ MeV and } \gamma_{5/2} = 0.37 \text{ fm}^{-2}. \tag{2}$$

This potential leads to resonance at 1.41(1) MeV (l.s.) and a width $\Gamma_{c.m.}$ of 100(1) keV that is in a complete agreement with the data [29]. Its phase shift is shown in Fig. 5 by the red dashed line.

The potential of the $^2P_{3/2}$ resonance state at an energy of 2.654 MeV (c.m.) with width 206(8) keV (c.m.) relative to the GS or 2.37 MeV (l.s.) above the threshold of the $n^{10}$Be channel without the FS has the following parameters:

$$V_{P3/2} = 10935.65 \text{ MeV and } \gamma_{1/2} = 40.0 \text{ fm}^{-2}. \tag{3}$$

This leads to a resonance energy of 2.37(1) MeV (l.s.) at a width of 204(1) keV (c.m.), and its phase shift is shown in Fig. 5 by the dashed blue line.

The potential of the first $^2P_{1/2}$ excited state of $^{11}$Be without FS has the following parameters:

$$V_{P1/2} = 192.799497 \text{ MeV and } \gamma_{1/2} = 0.7 \text{ fm}^{-2}. \tag{4}$$

This potential leads to a binding energy of $-0.181560$ MeV at an FDM accuracy of $10^{-6}$ MeV [42]. The AC is 0.27(1) in the range 10–30 fm, the mass radius is 2.61 fm, and the charge radius is 2.40 fm. This potential phase shift is shown in Fig. 5 by the red dotted/dashed line. The parameters of potential (4) have been optimized to correctly describe the total cross sections of the neutron capture on $^{10}$Be at a thermal energy of 25.3 meV, as obtained in [4], and the value of its dimensionless AC is relatively near to the lower limit of 0.29–0.71.

The other parameters of this potential are presented in [9]. In this work, the matrix elements (ME) for the computation of the total cross sections were calculated at distances of up to 30 fm, and the parameters of this potential were also taken from [4] to correctly describe the neutron capture on $^{10}$Be total cross sections at a thermal energy of 25.3 meV. It was subsequently ascertained that due to the extremely small value of the binding energy for the GS, and particularly for the FES, the ME calculation should be performed at large distances. Table 5 shows the values of the total cross sections for the GS and FES capture, and their convergence depending on the upper integration limit $R_{max}$ in ME.

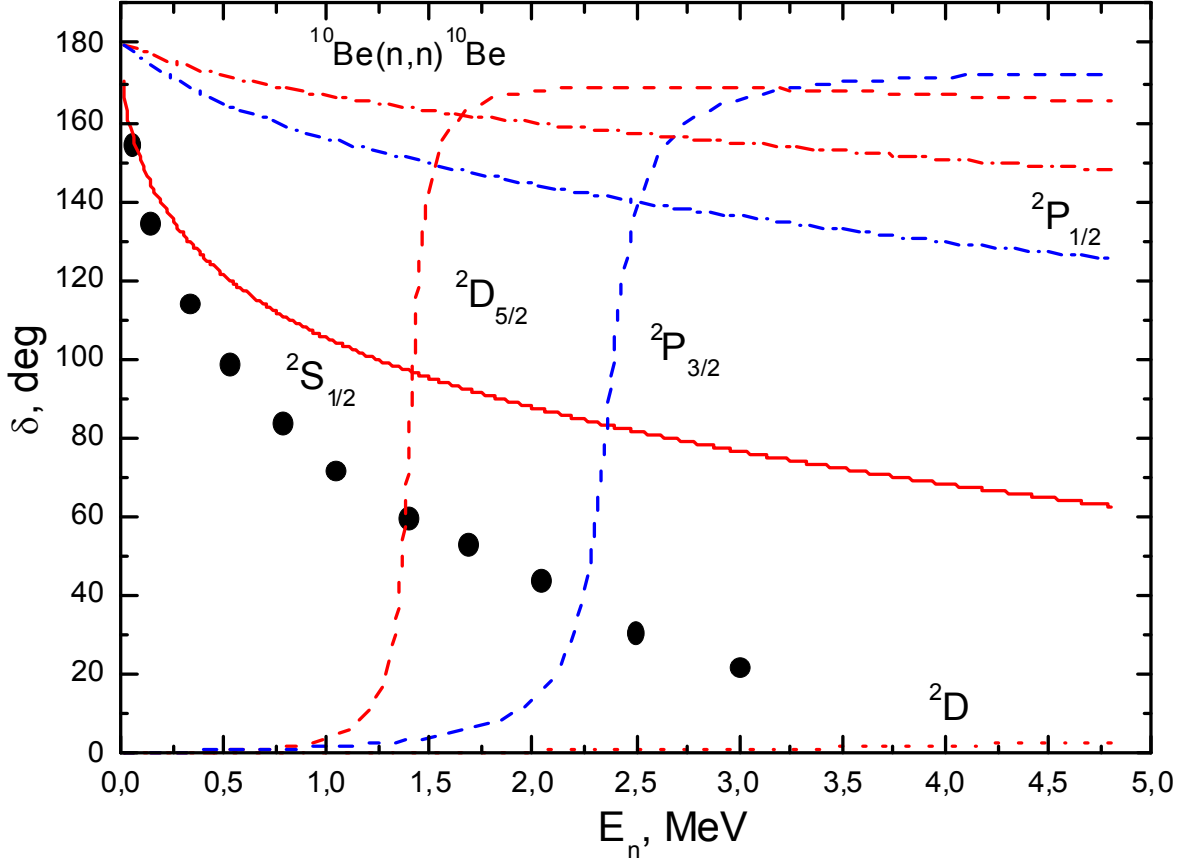

**Fig. 5.** $n^{10}$Be elastic scattering phase shifts in $^2S_{1/2}$ and $^2D$ waves: the solid red line shows the phase shift of potential (1); the dashed red line shows the phase shift of potential (2); the dashed blue line shows the phase shift of potential (3); the dotted red /dashed line shows the phase shift of potential (4); the dotted-dashed blue line shows the phase shift of potential (5); and the dotted red line shows the nonresonance $^2D$ scattering phase shift without spin-orbital splitting. Black points show $n^{10}$Be scattering $^2S_{1/2}$ phase shifts obtained in [43].

**Table 5.** Value of the neutron capture on $^{10}$Be total cross section in the transition to the GS for potential (1) and to the FES for potential (4), calculated at 1 keV

| $R_{max}$, fm | GS $\sigma_{tot}$, μb | FES $\sigma_{tot}$, μb |
|---|---|---|
| 30 | 0.27 | 0.21 |
| 50 | 0.56 | 0.92 |
| 100 | 0.64 | 1.50 |
| 150 | 0.64 | 1.52 |
| 200 | 0.64 | 1.52 |

As can be seen from Table 5, the ME have the correct values only at distances of 100-150 fm. Hence, in all our calculations of the total cross sections of this system we used an ME integration distance of 150 fm. As a result, new parameters for potential (4) were obtained. The GS potential was also changed due to a change in the capture total cross sections presented in Section 3.5, and the parameters of this potential were therefore used for the correct description of new data in the energy region of the order of 1 MeV. In addition, a new potential (3) is considered in the current work that enables to consider the resonance at the second transition No.1, which was not taken into account in previous work [9].

The results reported in [9] can be considered as taking into account only the core effects in $^{11}$Be. However, in the present work, we have taken into account the halo effects caused by the odd neutron of this nucleus. These effects appear generally in the cross-sections of capture to the FES. As can be seen from the tables given above, when $R_{max}$ increases from 30 to 150 fm, the total cross-section increases by about 2.5 times for capture to the GS, and by almost six times for the capture to the FES. Since the $^2P_{1/2}$ potential of the FES of $^{11}$Be is constructed in order to correctly describe the thermal cross-sections, the difference between the results obtained here and those reported in [9] at low energies lies only in its parameters. The results for the total cross sections obtained in the present work using new parameters (4) differ only slightly from the results obtained in [9] using the previous parameters. Furthermore, the capture cross section for transitions to the GS is increased a factor of two at low energies, since in this case the ME is calculated up to a distance of 150 fm.

We now return to the consideration of the criteria for construction of the $^2P_{1/2}$ scattering wave potential, which may differ from the FES potential due to the difference between the Young tableaux for these states [2]. Firstly, as shown above, this potential should not involve the FS. As we do not have the results of the phase shift analysis for the $n^{10}$Be elastic scattering, and since in the $^{11}$Be spectra there is no resonance with an accurately determined moment $J^\pi = 1/2^-$ and a known width [29] at an energy lower than 10.0 MeV, we assume that the $^2P_{1/2}$ potential must give rise to scattering phase shifts of almost zero in this energy region, and therefore its parameters can have zero width. For the $^2S_{1/2}$ scattering potential, the interaction between the GS and the FS for $^2S_{1/2}$, i.e. the parameters of potential (1), will be used.

## 4. Results

In this section, we discuss the results for the neutron radiative capture on $^{10}$Be total cross sections and the reaction rate, which can be obtained on the basis of the nuclear model and the potentials given above.

### *4.1. Total cross section*

As mentioned previously, we assume that radiative E1 capture No.1 occurs from the $^2P$ scattering waves to the $^2S_{1/2}$ GS of $^{11}$Be in the $n^{10}$Be channel. Our calculation of the capture cross sections for the GS potential (1) gives the results shown in Fig. 6 by the blue solid line. In all our calculations for the $^2P_{1/2}$ elastic scattering potentials, we used a potential of zero width, and for $^2P_{3/2}$ scattering potentials the potential with parameters (3) was used. The experimental data for neutron radiative capture on $^{10}$Be shown in Fig. 6 were recalculated here based on data in [11-13]. The results show that the calculated cross-sections generally describe the experimental data recalculated here [11,13] in the energy region under consideration. These results appear to be in good agreement with the results reported in [13], and do not exceed the limits of available experimental errors of this work.

Results reported in other works and our previous calculation results are presented below for comparison. The blue dashed line in Fig. 6 shows the results from [10], and the black dashed/dotted line represents the results reported in [4]. The green dashed line represents our previous results [9]. It should be borne in mind that these results were obtained with zero potentials for both $P$ waves; that is, the resonance in the $^2P_{3/2}$ scattering wave was not taken into account. The black solid line shows the present calculation results obtained by using the potentials from [9] while the ME integrating up to 150 fm. In contrast to the results obtained by the authors in [9], these results closely reproduce the data from [10] at all energies.

As can be seen from Fig. 6, the calculated cross section (solid blue line) decreases slowly at energies on the order of 10 keV and lower; this does not enable to calculate its value at a thermal energy of 290(90) μb, as presented in [4]. We therefore consider the *E*1 transitions from $^2S_{1/2}$ and $^2D_{3/2}$ scattering waves to the FES of $^2P_{1/2}$, which were given above as transition No.2. The blue dashed line in Fig. 7 presents the results for the *E*1 transition No.1 to the GS with potentials (1) and (3), and a zero potential for $^2P_{1/2}$ scattering. These results are shown in Fig. 6 by the solid blue line. The black dotted line corresponds to the *E*1 transition No.2 from the $^2S_{1/2}$ and $^2D_{3/2}$ scattering waves with potential (1) and moments *L* of 0 and 2, to the FES of $^2P_{1/2}$ with potential (4). The solid blue line shows the total cross section for these two E1 processes and correctly reproduces the general trend of the available experimental data over the whole energy region under consideration, from a thermal energy of 25.3 meV to an energy of about 5.0 MeV. The cross section calculated at the thermal energy takes a value of 302 μb. The experimental data at the thermal energy were taken from [4] - blue triangle with value equal to 290(90) μb, and [44] - black square corresponding to the upper limit of the thermal cross-section of 1 mb.

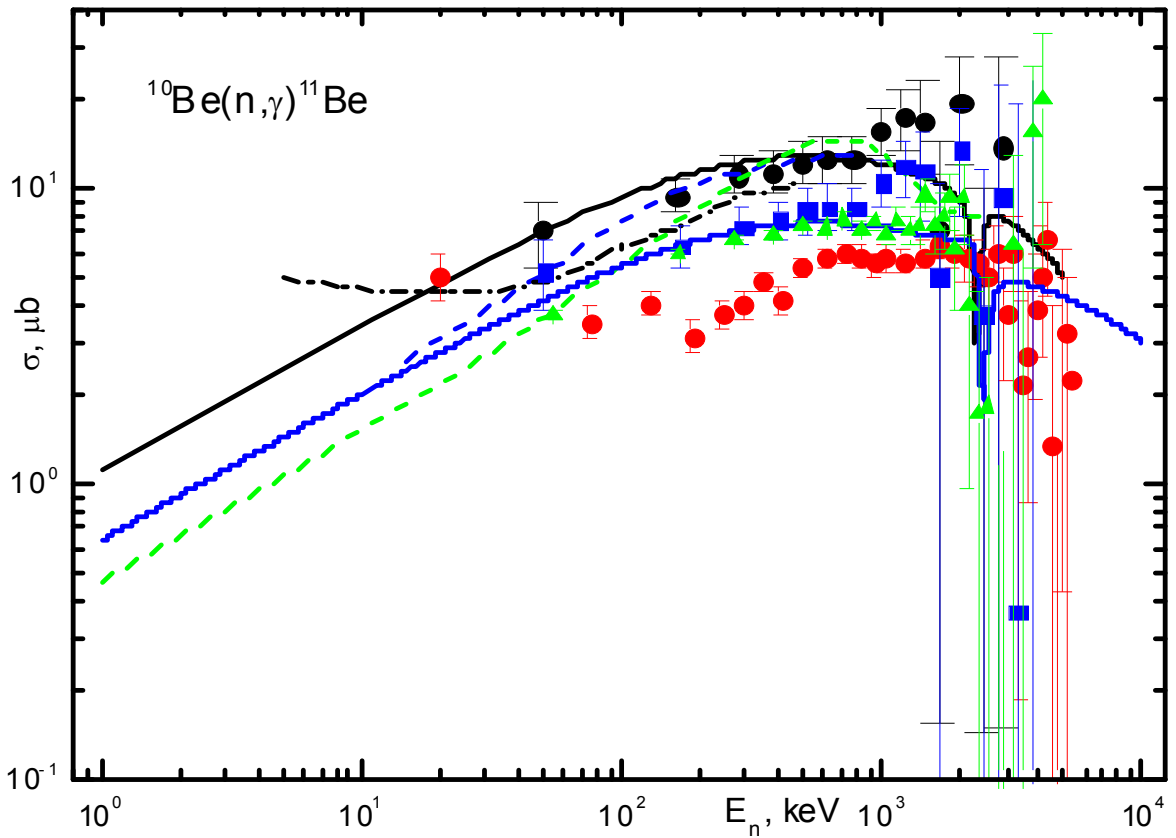

**Fig. 6.** Total cross sections of $^{10}$Be(*n*,γ)$^{11}$Be radiative *E*1 capture to the GS. Black points correspond to data from [10], based on the results in [11]; red points show the present recalculation of data [12]; green triangles show the recalculation of data from [13], blue squares are the present recalculation of data from [11]. The dashed blue line shows the results from [10]; the dotted/dashed black line represents the results from [4]; the dashed green line shows our previous results from [9]; the solid black line shows the results of the present calculation with potentials from [9]; and the solid blue line shows the results of present calculations with potentials (1), (3), and zero potential for $^2P_{1/2}$ scattering.

The cross-section of the allowed *M*1 transition from the $^2S_{1/2}$ scattering wave to the $^2S_{1/2}$ GS of $^{11}$Be in the $n^{10}$Be channel with the same potential (1) in both states will tend to zero due to the orthogonality of the discrete and continuous spectra wave functions in one potential. The real numerical calculation of such cross sections leads to a value that is lower than $10^{-2}$ μb in an energy region from 1 keV to 3.0 MeV. At an energy of 25.3 meV, the difference between this cross section and the cross section of transition to the FES is less than 1%, as shown in Fig. 7 by the black dotted line. If we consider the *M*1 transitions from the $^2P$ scattering waves with zero potential for the $^2P_{1/2}$ wave and potential (3) for the $^2P_{3/2}$ waves to the $^2P_{1/2}$ FES with potential (4), then the cross sections do not exceed a value of 0.15 μb in all energy ranges. For the $E2$ transitions from the $^2D_{3/2}$ wave with potential (1) at $L = 2$ and the $^2D_{5/2}$ wave with potential (2) to the GS with $^2S_{1/2}$, the value of the cross-section does not exceed $10^{-3}$ μb even at the resonance energies. It can therefore be seen that these transitions do not make a noticeable contribution to the total cross section of the process under consideration.

Fig. 7 illustrates the difference between the total cross sections at the thermal energy. In [4], a value of 290(90) μb was obtained, and in [44], the upper value was 1 mb, i.e. three times higher. Another potential for the FES may therefore be put forward, for example with the following parameters:

$V_{P1/2} = 42.112565$ MeV and $\gamma_{1/2} = 0.15$ fm$^{-2}$. (5)

This potential leads to a binding energy of −0.181560 MeV at an FDM accuracy of $10^{-6}$ MeV [42]. The AC is 0.40(1) in the range 7–30 fm, the mass radius is 2.90 fm, and the charge radius is 2.43 fm. This potential phase shift is illustrated in Fig. 5 by the blue dotted/dashed line, and the AC value is within a range of possible values from 0.29 to 0.81. The total cross sections with this potential are shown in Fig. 7 by the red dotted line, and at thermal energy this leads to cross sections of 809 μb. Hence, the significant ambiguity in the thermal cross sections and a wide range for the possible values of AC mean that we are unable to uniquely determine the potential parameters of the FES.

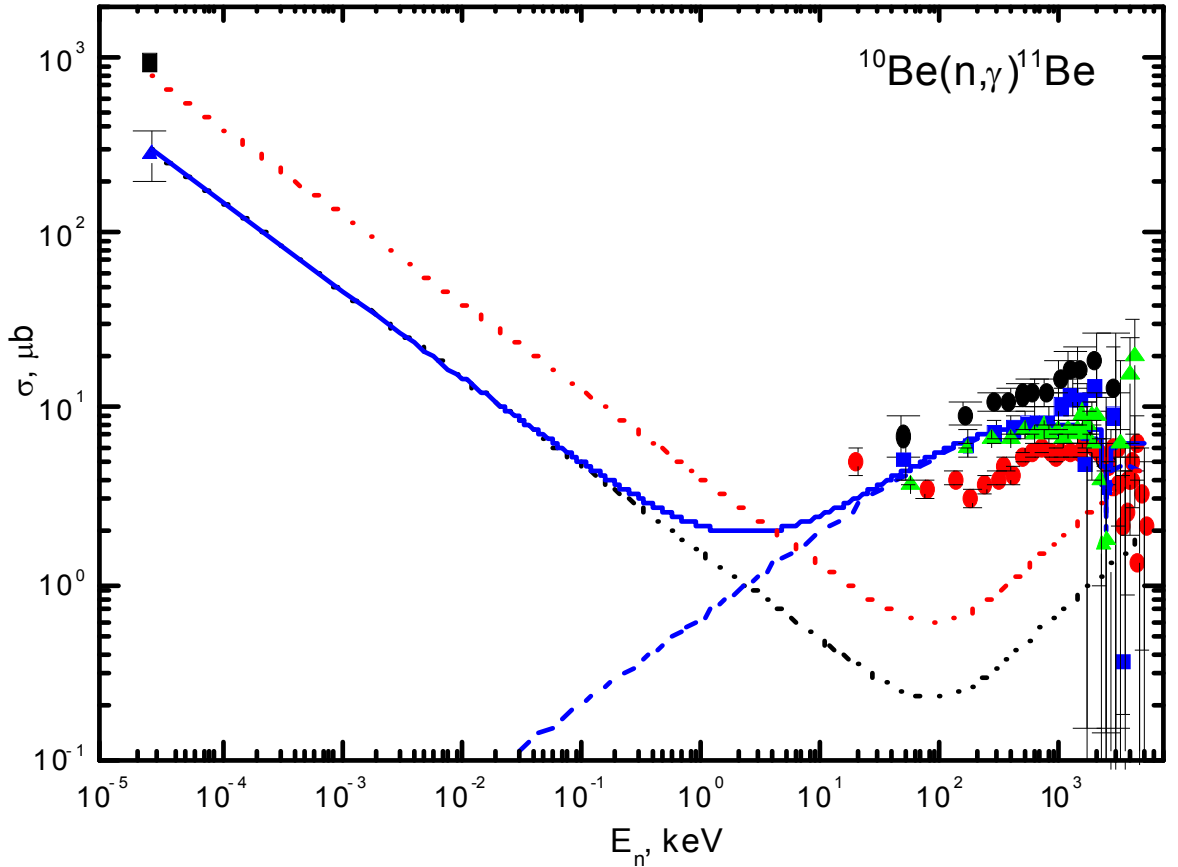

**Fig. 7.** Total cross-sections of the $^{10}$Be$(n,\gamma)^{11}$Be radiative capture. Black points represent data from [10], based on the results in [11]; red points correspond to the recalculation of data from [12]; green triangles show the recalculation of data from [13]; blue squares correspond to the present recalculation of the data in [11]; blue triangles represent our recalculation of the data in [4]; and black squares show the present recalculation of data in [44]. The solid blue line represents the present calculation; the dotted black line shows the cross sections of the transition to the FES (the present calculation); the dotted red line shows the present calculation with potential (5); and the dashed blue line shows the present calculations with potentials (1), (3), and zero potential for the $^2P_{1/2}$ scattering.

Since at energies of between 25.3 meV and about 10 eV the calculated cross section is a straight line (the blue solid line in Fig. 7), it can be approximated by a simple energy function of the following form:

$$\sigma_{ap}(\mu b) = \frac{A}{\sqrt{E(\text{keV})}} \ . \qquad (6)$$

The constant value $A = 1.5218$ μb·keV$^{1/2}$ was determined based on the point of minimum energy in the calculated cross sections of 25.3 meV. The modulus of relative deviation

$$M(E) = \left| [\sigma_{ap}(E) - \sigma_{theor}(E)] / \sigma_{theor}(E) \right|$$

of the calculated theoretical cross-section ($\sigma_{theor}$) and the approximation ($\sigma_{ap}$) of this cross section by the function in Equation (6) in an energy region of up to 10 eV is 0.2%. It is valid to assume that this form of the total cross-sectional dependence on energy will also be the same at lower energies. Therefore, on the basis of the expression given for the approximation of the cross section in Equation (6), we can evaluate the cross section at an energy of 1 μeV (1 μeV = $10^{-6}$ eV = $10^{-9}$ keV), for example, that gives a value on the order of 48.1 mb.

*4.2. Reaction rates*

Fig. 8 shows the reaction rate $N_A\langle\sigma v\rangle$ of the neutron radiative capture on $^{10}$Be (solid red line) in the region 0.01–10.0 $T_9$. This corresponds to the solid blue line in Fig. 7, and is presented in the usual form [17]:

$$N_A\langle\sigma v\rangle = 3.7313 \cdot 10^4 \mu^{-1/2} T_9^{-3/2} \int_0^\infty \sigma(E) E \exp(-11.605 E / T_9) dE \ ,$$

where E is given in MeV, the cross-section σ(E) is measured in μb, μ is the reduced mass in amu, and $T_9$ is the temperature in $10^9$ K. To calculate this rate, the total cross-section shown in Fig. 7 was computed in the region 10.0 meV to 10.0 MeV. The dotted/dashed blue line in Fig. 8 shows the reaction rate of capture to the GS, and the dotted/dashed red line shows the rate of capture to the FES.

The solid black line presents the results from [4] for the approximation of the calculated rate of the reaction under consideration. This curve is appreciably higher than the present results at temperatures of less than 1.0 $T_9$. This is explained by using in [4] the cross sections agreed with data [10]. In Fig. 8, the solid green line shows the results of the reaction rate from [45], which also agree with the results for the cross-sections from [10]. The reaction rate from [46] is shown by the solid blue line; it has a very different form and shows the least agreement with the results given above.

The solid red line in Fig. 8 in the region 0.01–10.0 $T_9$ can be approximated by an expression of the form

$$N_A\langle\sigma v\rangle = \sum_{k=1}^{6} a_k T_9^{k-1} \qquad (7)$$

with the parameters from Table 6.

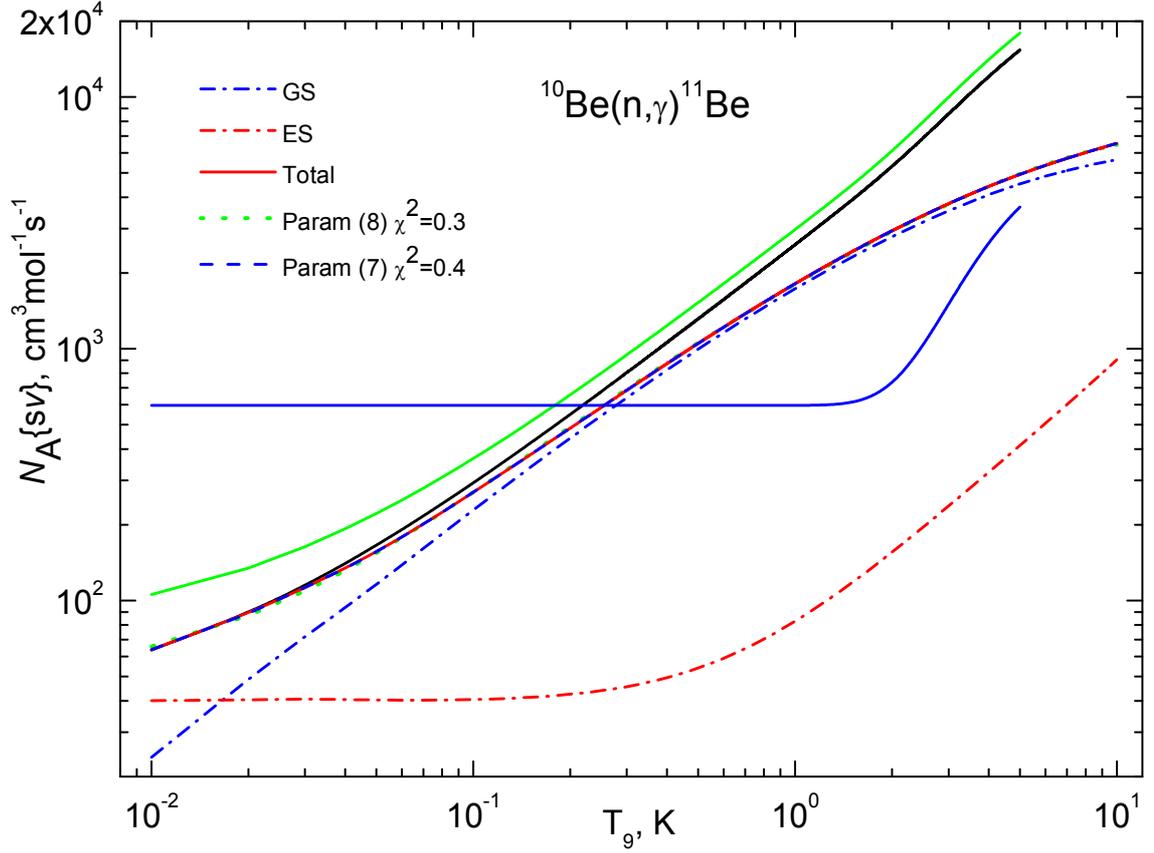

**Fig. 8.** Reaction rate of the neutron capture on $^{10}$Be. The dotted/dashed blue line represents the reaction rate for capture to the ground state (present calculation); the dotted/dashed red line is the reaction rate for capture to the first excited state (present calculation); the solid red line shows the total reaction rate (present calculation); the dashed blue line shows the approximation of the calculated reaction rate at $\chi^2 = 0.4$ (7); the dotted green line shows the approximation of the calculated reaction rate at $\chi^2 = 0.3$ (8); the solid green line shows the results from [45]; the solid blue line shows the results from [46]; and the solid black line shows the results of approximation of the calculated reaction rate from [4].

**Table 6.** Extrapolation parameters for Equation 7

| $k$ | 1 | 2 | 3 | 4 | 5 | 6 |
|---|---|---|---|---|---|---|
| $a_k$ | 44.23061 | 2271.351 | −580.9702 | 105.1334 | −9.97608 | 0.36633 |

The results of approximation with these parameters are shown in Fig. 8 by the dashed blue line at an average value of $\chi^2 = 0.4$ and 1% errors of the theoretical reaction rate. Another form of approximation of the calculated reaction rate [47] can also be used:

$$N_A\langle\sigma v\rangle = 5.9688/T_9^{2/3} \cdot \exp(-0.34181/T_9^{1/3}) \cdot (1.0 + 0.01441 \cdot T_9^{1/3} + 55.30271 \cdot T_9^{2/3} - 287.1127 \cdot T_9 + 883.6600 \cdot T_9^{4/3} - 227.0900 \cdot T_9^{5/3}) \quad (8)$$

With $T_9 = 10^9$ K, this also gives $\chi^2 = 0.3$ at 1% errors of the calculated reaction rate. The results of this approximation are shown in Fig. 8 by the dotted green line.

**Conclusions**

Using the framework of the MPCM with a classification of states based on the Young tableaux, we have succeeded in constructing the potentials for $n^{10}$Be interaction, which enables us to correctly reproduce the general trend of the available experimental data for the total cross sections of neutron radiative capture on $^{10}$Be at low and ultralow energies. These theoretical cross sections are calculated from the thermal energy, in the range 10.0 meV to 10.0 MeV, and are approximated by a simple function for energy (shown in Eq. (6)) that can also be used for the calculation of the cross sections at energies of less than 10 eV. The reaction rate is calculated, and two forms of its approximation are presented for temperatures of between 0.01 and 10.0 $T_9$, for a small value of $\chi^2$.

The proposed variants of the potentials of the ground and first excited states of $^{11}$Be in the $n^{10}$Be channel enable us to obtain the AC in the limits of errors available for it, and lead to a reasonable description of the $^{11}$Be radii. These potentials are then used to calculate the characteristics of the thermonuclear neutron radiative capture reaction on $^{10}$Be. The results of this calculation are in good agreement with the present results for the total radiative capture cross sections from the data [11,13] for $^{11}$Be Coulomb dissociation probability. On the basis of the results obtained above, we suggest the form of Equation (8) for the reaction rate approximation shown in Fig. 8; this appears to be the best fit for the rate description at minimal $\chi^2$.

**Acknowledgments**

The authors would like to express their deep gratitude to Prof. T. Aumann and Prof. T. Nakamura for providing the tabulated data on the Coulomb dissociation probability, published earlier in [12,16] and [13], respectively. We also very much appreciate the important remarks of Prof. T. Aumann on this manuscript.

This work was supported by the Ministry of Education and Science of the Republic of Kazakhstan, grant no. BR05236322, entitled "Study reactions of thermonuclear processes in extragalactic and galactic objects and their sybsystems" through the Fesenkov Astrophysical Institute of the National Center for Space Research and Technology of the Ministry of Defense and Aerospace Industry of the Republic of Kazakhstan (RK).

**References**

[1]   S.B. Dubovichenko, Radiative neutron capture and primordial nucleosynthesis of the Universe, fifth Russian ed., corrected and added, Lambert Academy Publ. GmbH&Co. KG., Saarbrucken, 2016; S.B. Dubovichenko, Radiative neutron capture and primordial nucleosynthesis of the Universe, first English ed., Springer, London, 2018 (in Press).

[2]   O.F. Nemets, V.G. Neudatchin, A.T. Rudchik, Y.F. Smirnov, Yu.M. Tchuvil'sky, Nucleon association in atomic nuclei and nuclear reactions of many nucleons transfers, Naukova dumka, Kiev, 1988 (in Russian).

[3]   S.B. Dubovichenko, Thermonuclear processes in Stars and Universe, second English edition, Scholar's Press, Saarbrucken, 2015; https://www.scholars-press.com/catalog/details/store/gb/book/978-3-639-76478-9/Thermonuclear-processes-in-stars.

[4]   Liu Zu-Hua, Zhou Hong-Yu, Nuclear halo effect on nucleon capture reaction rate at stellar energies, Chinese Phys. 14 (2005) 1544-1548.

[5]   V. Guimaraes, et al., Investigation of nucleosynthesis neutron capture reactions using transfer reactions induced by $^8$Li beam, in: Proceedings of the International Symposium on Nuclear Astrophysics - Nuclei in the Cosmos. IX. June 25-30. CERN, Geneva, Switzerland, 2006, p.108.

[6]   M. Terasawa, New nuclear reaction flow during r-process nucleosynthesis in supernovae: critical role of light, neutron-rich nuclei, Astrophys. J. 562 (2001) 470-479.

[7]   Toshitaka Kajino, et al., Fusion Reactions in Supernovae and the Early Universe, Prog. Theor. Phys. Supp. 154 (2004) 301.

[8] T. Sasaqui, T. Kajino, G.J. Mathews, K. Otsuki, T. Nakamura, Sensitivity of *r*-process nucleosynthesis to light-element nuclear reactions, Astrophys. J. 634 (2005) 1173-1189.

[9] S.B. Dubovichenko, A.V. Dzhazairov-Kakhramanov, The radiative $^{10}$Be(n,γ)$^{11}$Be capture at thermal and astrophysical energies, J. Phys. G 43 (2016) 095201(14p.).

[10] A. Mengoni, et al., Exotic properties of light nuclei and their neutron capture cross sections, in: Proceedings of the 4$^{th}$ International Seminar on Interaction of Neutrons with Nuclei "Neutron Spectroscopy, Nuclear Structure, Related Topics" Dubna (Russia), April 1996; arXiv:nucl-th/9607023 [nucl-th].

[11] T. Nakamura, et al., Coulomb dissociation of a halo nucleus $^{11}$Be at 72 MeV, Phys. Lett. B 331 (1994) 296-301.

[12] R. Palit, et al., Exclusive measurement of breakup reactions with the one-neutron halo nucleus $^{11}$Be, Phys. Rev. C 68 (2003) 034318(1-14).

[13] S. Nakamura, Y. Kondo, Neutron Halo and Breakup Reactions, Clusters in Nuclei, Ed. by C. Beck, V.2., Lecture Notes in Physics, 848 (2012) pp. 67-119.

[14] https://physics.nist.gov/cgi-bin/cuu/Value?alph|search_for=atomnuc!

[15] http://cdfe.sinp.msu.ru/services/ground/NuclChart_release.html .

[16] T. Aumann, private commun., 2017.

[17] C. Angulo, et al., A compilation of charged-particle induced thermonuclear reaction rates, Nucl. Phys. A 656 (1999) 3-183.

[18] S.B. Dubovichenko, A.V. Dzhazairov-Kakhramanov, The reaction $^{8}$Li(n,γ)$^{9}$Li at astrophysical energies and its role in primordial nucleosynthesis, Astrophys. J. 819 (2016) 78(8p.).

[19] S.B. Dubovichenko, A.V. Dzhazairov-Kakhramanov, Study of the Nucleon Radiative Captures $^{8}$Li(n,γ), $^{9}$Be(p,γ), $^{10}$Be(n,γ), $^{10}$B(p,γ), and $^{16}$O(p,γ) at Thermal and Astrophysical Energies, Int. Jour. Mod. Phys. E 26 (2017) 1630009(56p.).

[20] E.G. Adelberger, et al., Solar fusion cross sections II. The pp chain and CNO cycles, Rev. Mod. Phys. 83 (2011) 195-245.

[21] V.I. Kukulin, V.G. Neudatchin, I.T. Obukhovsky and Yu.F. Smirnov, Clusters as subsystems in light nuclei, in: K. Wildermuth and P. Kramer (Eds.) Clustering Phenomena in Nuclei, Vieweg, Branschweig, 1983, V.3, p.1.

[22] S.B. Dubovichenko, Yu.N. Uzikov, Astrophysical *S*-factors of reactions with light nuclei, Phys. Part. Nucl. 42 (2011) 251-301.

[23] A.M. Mukhamedzhanov, R.E. Tribble, Connection between asymptotic normalization coefficients, sub threshold bound states, and resonances // Phys. Rev. C 59 (1999) 3418-3424.

[24] L.D. Blokhintsev, I. Borbey, E.I. Dolinsky Nuclear vertex constants, Phys. Part. Nucl. 8 (1977) 1189-1245.

[25] G.R. Plattner, R.D. Viollier, Coupling constants of commonly used nuclear probes, Nucl. Phys. A 365 (1981) 8-12.

[26] V.G. Neudatchin, Yu.F. Smirnov, Nucleon associations in light nuclei, Nauka, Moscow, 1969 (in Russian).

[27] C. Itzykson, M. Nauenberg, Unitary groups: Representations and decompositions, Rev. Mod. Phys. 38 (1966) 95-101.

[28] D.R. Tilley, et al., Energy level of light nuclei. A = 8,9,10, Nucl. Phys. A 745 (2004) 155-363.

[29] J.H. Kelley, et al., Energy level of light nuclei A = 11, Nucl. Phys. A 880 (2012) 88-195.

[30] S.B. Dubovichenko, A.V. Dzhazairov-Kakhramanov, Thermonuclear processes for three body system in the potential cluster model, Nucl. Phys. A 941 (2015) 335-363.

[31] S.B. Dubovichenko, A.V. Dzhazairov-Kakhramanov, Neutron radiative capture on $^{10}$B, $^{11}$B and proton radiative capture on $^{11}$B, $^{14}$C and $^{15}$N at thermal and astrophysical energies, Int. Jour. Mod. Phys. E 23 (2014) 1430012(1-55).

[32] S.B. Dubovichenko, A.V. Dzhazairov-Kakhramanov, N.V. Afanasyeva, New Results for

Reaction Rate of the Proton Radiative Capture on $^3$H, Nucl. Phys. A 963 (2017) 52-67.

[33] V.G. Neudatchin, et al., Generalized potential model description of mutual scattering of the lightest p$^2$H, $^2$H$^3$He nuclei and the corresponding photonuclear reactions, Phys. Rev. C 45 (1992) 1512-1527.

[34] W. Nortershauser, et al., Nuclear Charge Radii of $^{7,9,10}$Be and the One-Neutron Halo Nucleus $^{11}$Be, Phys. Rev. Lett. 102 (2009) 062503.

[35] H.W. Hammer, D.R. Phillips, Electric properties of the Beryllium-11 system in Halo EFT, Nucl. Phys. A 865 (2011) 17-42.

[36] T.L. Belyaeva, et al., Neutron asymptotic normalization coefficients and halo radii of the first excited states of $^{12}$C and $^{11}$Be, EPJ Web of Conferences 66 (2014) 03009.

[37] T. Aumann, et al., One-Neutron Knockout from Individual Single-Particle States of $^{11}$Be, Phys. Rev. Lett. 84 (2000) 35.

[38] N. Fukuda, et al., Coulomb and nuclear breakup of a halo nucleus $^{11}$Be, Phys. Rev. C 70 (2004) 054606(1-12).

[39] S.I. Sukhoruchkin, Z.N. Soroko, Exited nuclear states. Sub.G. Suppl. I/25 A-F. Springer. 2016.

[40] D.L. Auton, Direct reactions on $^{10}$Be, Nucl. Phys. A 157 (1970) 305-322; D.R. Goosman, R.W. Kavanagh, $^{10}$Be(d,p)$^{11}$Be and the $^{10}$Be(d,α)$^8$Li Reactions, Phys. Rev. C. 1 (1970) 1939.

[41] B. Zwieglinski, et al., Study of the $^{10}$Be(d, p)$^{11}$Be reaction at 25 MeV, Nucl. Phys. A 315 (1979) 124-132.

[42] S.B. Dubovichenko, Calculation method of the nuclear characteristics, Complex, Almaty, 2006; arXiv:1006.4947 [nucl-th], (in Russian).

[43] S. Quaglioni, P. Navrátil, Ab initio many-body calculations of nucleon-nucleus scattering, Phys. Rev. C 79 (2009) 044606(1-28).

[44] S.F. Mughabghab, Atlas of neutron resonances, National Nuclear Data Center, Brookhaven, National Laboratory, Upton, USA, 2006, 1008p.

[45] A. Mengoni, et al., Exotic structure of light nuclei and their neutron capture reaction rates, Nucl. Phys. A 621 (1997) 323-326.

[46] T. Rauscher, et al., Production of heavy elements in inhomogeneous cosmologies, Astrophys. J. 429 (1994) 499-530.

[47] G.R. Caughlan, W.A. Fowler, Atom. Data Nucl. Data Tab. 40 (1988) 283-334.